\def\vec#1{{\rm\bf #1}}

\documentclass[11pt]{article} 
\usepackage{graphics}
\usepackage[dvips]{graphicx}
\usepackage{amssymb}
\usepackage{amsmath}
\usepackage{vmargin}
\begin{document}

\title{The missing stress-geometry equation in granular media}
\author{S. F. Edwards and D. V. Grinev \\
Cavendish Laboratory, University of Cambridge, Madingley Road,\\ Cambridge CB3 OHE, United Kingdom}

\maketitle

\begin{abstract}
The simplest solvable problem of stress transmission through a static granular material is when the grains are perfectly rigid 
and have an average coordination number of $\bar{z}=d+1$. Under these conditions there exists an analysis of stress which is independent 
of the analysis of strain and the $d$ equations of force balance $\nabla_{j}\, \sigma_{ij}({\vec r})\,=\,g_{i}({\vec r})$ have to be supported by $\frac{d(d-1)}{2}$ equations. These equations are of purely
geometric origin. A method of deriving them has been proposed in an earlier paper \cite{EG}. In this paper alternative derivations are discussed
and the problem of the "missing equations" is posed as a geometrical puzzle which has yet to find a systematic solution as against sensible but 
fundamentally arbitrary approaches.
\end{abstract}

\section{Introduction}
\begin{figure}[h] 
\begin{center} 
\resizebox{8cm}{!}{\rotatebox{0}{\includegraphics{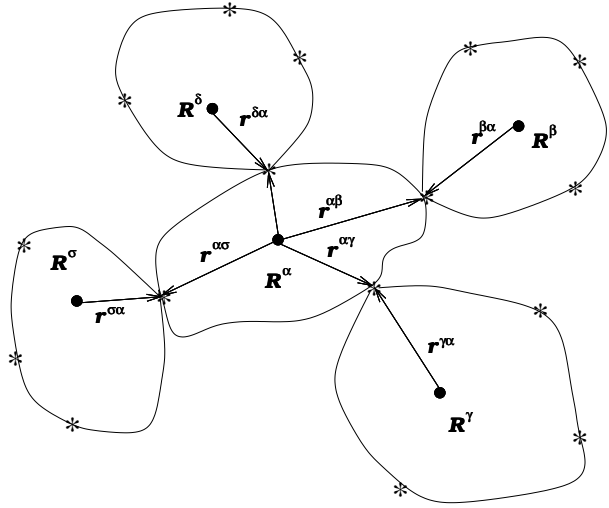}}}
\caption{Cross-section of the first coordination shell of the reference particle $\alpha$}
\label{shell} 
\end{center} 
\end{figure}

Granular media are ubiquitous yet complex materials with puzzling properties \cite{Jaeger}. 
The simplest model of a static granular material is that where grains are considered to be perfectly hard, perefectly rough and each grain $\alpha$ has a 
coordination number $z^{\alpha}=d+1$ (where $d$ is the dimension of the system). Under these conditions Newton's equations of intergranular force and
couple balance can be solved \cite{EG}. The system is in the state of mechanical equilibrium and particles can not experience deformation under
load so there is no displacement field present. Thus the only immediately available macroscopic equation has the form  
    
\begin{equation}
\nabla_{j}\, \sigma_{ij}({\vec r})\,=\,g_{i}({\vec r})\, ,
\label{sfeqmacro}
\end{equation}
where $\sigma_{ij}({\vec r})$ is the macroscopic stress tensor and $g_{i}({\vec r})$ is external force at the boundaries. 
The vector equation (\ref{sfeqmacro}) gives $d$ equations for $\frac{d(d+1)}{2}$ components of $\sigma_{ij}({\vec r})$ leaving $\frac{d(d-1)}{2}$
further equations required to solve for the macroscopic stress tensor. Thus one should be able to derive them from the geometry of the contact network which 
is assumed to be specified. In order to put the above comments into formulae we draw a diagram (see Figure \ref{shell}) of one grain in 
contact with 4 nearest neighbours (i.e. grains that are in contact with the reference grain $\alpha$).

\begin{figure}[h] 
\begin{center} 
\resizebox{5cm}{!}{\rotatebox{0}{\includegraphics{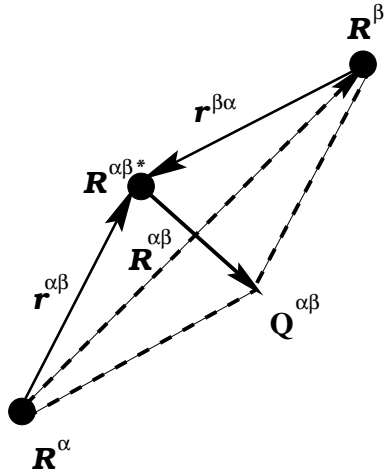}}}
\caption{Segment of the contact network between nearest neighbours $\alpha$ and $\beta$}
\label{centroid} 
\end{center} 
\end{figure}

The geometrical specification of the system is given by the set of $\frac{\sum_{\alpha}^{N}z^{\,\alpha}d}{2}$ contact points $\{{\vec{R}}^{\,\alpha\beta *}\}$. 
The centroid of contacts of the reference grain $\alpha$ is defined by vector  ${\vec R}^{\,\alpha}$

\begin{equation}
\vec{R}^{\,\alpha}=\frac{\sum_{\,\beta} \,\vec{R}^{\,\alpha\beta *}}{z^{\,\alpha}}\, ,
\label{conpoint0}
\end{equation}
where the summation sign $\sum_{\beta}$ means the sum over all nearest neighbours of the grain 
$\alpha$. The distance between grains $\alpha$ and $\beta$ is defined as the distance between their centroids of contacts 
 
\begin{equation}
\vec{R}^{\,\alpha\beta}=\vec{R}^{\,\beta}- \vec{R}^{\,\alpha}\,=\vec{r}^{\,\alpha\beta}- \vec{r}^{\,\beta\alpha} ,
\label{centroiddis0}
\end{equation}
where $\vec{r}^{\,\alpha\beta}$ is the vector joining the centroid of contact with the contact point. The second vector which characterises the 
relative position of neighbouring centroid with respect to the contact point is defined by (see Figure \ref{centroid})

\begin{equation}
\vec{Q}^{\,\alpha\beta}\,=\,-(\vec{r}^{\,\alpha\beta} \,+\, \vec{r}^{\,\beta\alpha}) \, .
\label{qvec}
\end{equation}

\begin{figure}[h] 
\begin{center} 
\resizebox{6cm}{!}{\rotatebox{0}{\includegraphics{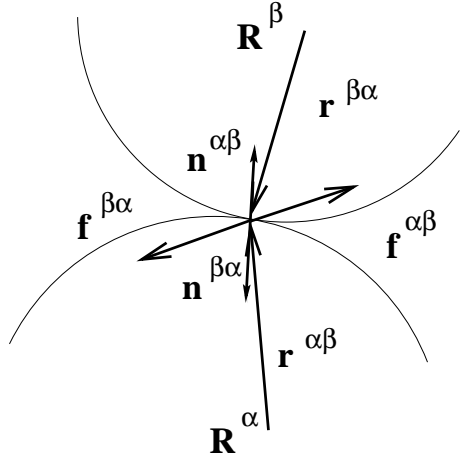}}}
\caption{Intergranular forces and the local geometry of two grains in contact}
\label{2grains} 
\end{center} 
\end{figure}

Newton's laws of force and couple balance for every grain give us the system of $\frac{Nd(d+1)}{2}$ equations for $\frac{zdN}{2}$ 
interparticle forces ${\vec f}^{\,\alpha\beta}$ (see Figure \ref{2grains})

\begin{equation}
\sum_{\beta}f^{\,\alpha\beta}_{i}=g^{\alpha}_{i}\, ,
\label{newt2}
\end{equation}

\begin{equation}
f^{\,\alpha\beta}_{i}+f^{\,\beta\alpha}_{i}\,=\,0\, ,
\label{newt3}
\end{equation}

\begin{equation}
\sum_{\beta}\epsilon_{ikl}\,f^{\,\alpha\beta}_{k}\,r^{\,\alpha\beta}_{l}=c_{i}^{\,\alpha}\, .
\label{couple}
\end{equation}
where $g_{i}^{\,\alpha}$ is the external force acting on grain $\alpha$ and $c_{i}^{\,\alpha}$ is the external couple. 
The tensorial force moment $S^{\,\alpha}_{ij}$ for grain $\alpha$ is defined as
\begin{equation}
S^{\,\alpha}_{ij}=\sum_{\beta}\,f^{\,\alpha\beta}_{i}\,r^{\,\alpha\beta}_{j}\, ,
\label{microstress}
\end{equation}
and it is symmetric tensor given that $c_{i}^{\,\alpha}~=~0$.
The macroscopic stress tensor $\sigma_{ij}({\vec r})$ can be obtained by averaging  $S^{\,\alpha}_{ij}$ over the packing
\begin{equation}
\sigma_{ij}({\vec r})\,=\,\langle \,\,S_{ij}^{\,\alpha}\,\rangle\, ,
\label{macrostress}
\end{equation}
where $\langle \cdots \rangle = \frac{1}{V}\sum_{\alpha=1}^{N}\, \cdots \delta({\vec r}-{\vec R^{\,\alpha}})$ in the simplest case and $V$ is the volume of the packing.
The method offered by the authors in \cite{EG} was to consider the probability functional for the set $\{S^{\,\alpha}_{ij}\}$

\begin{equation}
\begin{split}
P \left \{ S_{ij}^{\,\alpha} \right \} &= {\mathcal M} \int\prod_{\alpha,\beta}
\delta\Big (S^{\,\alpha}_{ij}-\sum_{\beta}f^{\,\alpha\beta}_{i}r^{\,\alpha\beta}_{j}\Big ) 
\,P \left \{f^{\,\alpha\beta} \right \} \, {\mathcal D} f^{\,\alpha\beta}\, ,
\end{split}
\label{stresspdf}
\end{equation}
where
\begin{equation}
\begin{split}
 P \left \{f^{\,\alpha\beta} \right \} &= {\mathcal N} 
 \prod_{\alpha,\beta}\,\delta\big(\sum_{\beta}f^{\,\alpha\beta}_{i}-g^{\alpha}_{i}\big) \\
 &\mbox{}\times
\delta\big(\sum_{\beta}\epsilon_{ikl}\,f^{\,\alpha\beta}_{k}\,r^{\,\alpha\beta}_{l}\big) \\
 &\mbox{}\times
\delta\big(f^{\,\alpha\beta}_{i}+f^{\,\beta\alpha}_{i}\big)\, ,
\end{split}
\label{forcepdf}
\end{equation}
and the normalisations, ${\mathcal N}$ and  ${\mathcal M}$ are functions of contact network configuration.
The main goal of Ref.\cite{EG} was to transform (\ref{stresspdf}) into

\begin{equation}
\begin{split}
P \left \{ S_{ij}^{\,\alpha} \right \} \,=\, \prod_{\alpha}
\delta\Big (\sum_{\beta}K^{\,\alpha \beta}_{ijk}\,S^{\,\beta}_{jk}\,-\,g^{\,\alpha}_{i}\Big)\,
\delta\Big (\sum_{\beta}P^{\,\alpha \beta}_{ijkl}\,S^{\,\beta}_{kl}\Big)
\end{split}
\label{stresspdffinabs}
\end{equation}
where delta-functions contain the complete system of equations for the set of tensorial force moments $\{S^{\,\alpha}_{ij}\}$.
The method of Ref.\cite{EG} was to exponentiate all delta-functions in (\ref{stresspdf}) 
\begin{equation}
\begin{split}
P \left \{ S_{ij}^{\,\alpha} \right \}=\int \prod_{\alpha, \beta}^{N}\,e^{iA}\, {\mathcal D} f^{\,\alpha\beta}
\,{\mathcal D}{\vec \zeta}^{\,\alpha}\,{\mathcal D}{\vec \gamma}^{\,\alpha}\,{\mathcal D}
{\vec \lambda}^{\,\alpha}\,{\mathcal D}{\vec \eta}^{\,\alpha\beta}\, ,
\end{split}
\label{stresspdf2}
\end{equation}
where 
\begin{equation}
\begin{split}
A\,&=\,\sum_{\alpha}\,\zeta^{\,\alpha}_{ij}\Big (S^{\,\alpha}_{ij}-\sum_{\beta}f^{\,\alpha\beta}_{i}r^{\,\alpha\beta}_{j}\Big ) \\
&\mbox{}+
\gamma^{\,\alpha}_{i}\Big(\sum_{\beta}f^{\,\alpha\beta}_{i}-g^{\alpha}_{i}\Big) \\
&\mbox{}+
\lambda^{\,\alpha}_{i}\Big(\sum_{\beta}\epsilon_{ikl}f^{\,\alpha\beta}_{k}r^{\,\alpha\beta}_{l}\Big) \\
&\mbox{}+
\eta^{\,\alpha\beta}_{i}\Big(f^{\,\alpha\beta}_{i}\,+\,f^{\,\beta\alpha}_{i}\Big)\, . 
\end{split}
\label{entropy}
\end{equation}
After integrating out the fields $f_{i}^{\,\alpha\beta}$, $\lambda_{i}^{\,\alpha}$ and $\eta_{i}^{\,\alpha\beta}$  
the following system of linear equations for the conjugate fields $\{\zeta^{\,\alpha}_{ij}\}$ and $\{\gamma_{i}^{\,\alpha}\}$  was obtained

\begin{equation}
\zeta^{\,\alpha}_{ij}r^{\,\alpha\beta}_{j}-\gamma^{\,\alpha}_{i}=\zeta^{\,\beta}_{ij}r^{\,\beta\alpha}_{j}-\gamma^{\,\beta}_{i}\, .
\label{confieldeqn}
\end{equation}
From (\ref{confieldeqn}) it was shown in Ref.\cite{EG} that $\zeta^{\,\alpha}_{ij}$ has the representation

\begin{equation}
\zeta^{\,\alpha}_{ij}\,=\,\zeta^{\,\alpha\,0}_{ij}\,+\,\zeta^{\,\alpha\,*}_{ij}\, ,
\label{pscf}
\end{equation}
where a particular solution $\zeta^{\,\alpha\,0}_{ij}$  gave the first delta-function in (\ref{stresspdffinabs})
        
\begin{equation}
\sum_{\beta}\,S^{\,\alpha}_{ij}\,M^{\,\alpha}_{jl}\,R^{\,\alpha\beta}_{l}\,-\,\sum_{\beta}\,S^{\,\beta}_{ij}\,M^{\,\beta}_{jl}\,R^{\,\beta\alpha}_{l}\,=\,g^{\,\alpha}_{i}\, ,
\label{sfeqdis}
\end{equation}
and a complimentary function $\zeta^{\,\alpha\,*}_{ij}$  which satisfies the following system of linear equations 

\begin{equation}
\zeta^{\,\alpha\,*}_{ij}r^{\,\alpha\beta}_{j}\,-\,\zeta^{\,\beta\,*}_{ij}r^{\,\beta\alpha}_{j} = 0\, .
\label{zetastareq}
\end{equation}
This system of linear equations for $\{ \zeta^{\,\alpha\,*}_{ij} \}$ gave the required $\frac{Nd(d-1)}{2}$ constraints on  $\{ S^{\,\alpha}_{ij} \}$.
The present paper concentrates on $\zeta^{\,\alpha\,*}_{ij}$  and uses the result of Ref.\cite{EG} that

\begin{equation}
P \left \{S_{ij}^{\,\alpha} \right \}\,=\,\prod_{\alpha=1}^{N}\,
\delta \Big(\sum_{\beta}S^{\,\alpha}_{ij}M^{\,\alpha}_{jl} R^{\,\alpha\beta}_{l}-S^{\,\beta}_{ij}M^{\,\beta}_{jl}
R^{\,\beta\alpha}_{l} -g^{\,\alpha}_{i}\Big )\,P \left \{ S_{ij}^{\,\alpha}|geometry \right \}\, ,
\end{equation}
where $P \left \{ S_{ij}^{\,\alpha}|geometry \right \}$ contains the set of $\frac{d(d-1)}{2}$ missing equations which in continuum limit might take a form

\begin{equation}
 P_{ijkl}\sigma_{kl} + \nabla_{j}T_{ijkl}\sigma_{kl} + \nabla_{j}\nabla_{l}U_{ijkl}\sigma_{km} + ... =0 \, .
\label{sgeomeq}
\end{equation}
In the first term we have $ P_{ijkl}~=~ -P_{jikl}$ which gives the correct number of equations.
The missing equation (\ref{sgeomeq}) is of purely geometric origin. We deliberately avoid using the term "constitutive relation" (for there is no deformation or
displacement in this model) and call Eq. (\ref{sgeomeq}) the stress-geometry equation. The authors believe this kind of situation offeres a new kind of challenge in theoretical physics and
although this paper presents a solution, it does not have the elegance and completeness that one might expect from the solution derived by means of some variational principle.
There may be some analogy here to the problems of dynamics where, although the Lagrange equations with Lagrange multipliers will solve any non-holonomic problem, the really
powerful way is to use the Gibbs-Appell equations \cite{Pars}.

\section{The missing stress-geometry equation}

In order to obtain $P \left \{ S_{ij}^{\,\alpha}|geometry \right \}$ and derive  (\ref{sgeomeq}) we need to solve Eq.(\ref{zetastareq}). Let us describe the method of Ref.\cite{EG}. 
We notice that in order to obtain the precise number of missing equations (which is $\frac{Nd(d-1)}{2}$), Eq.(\ref{zetastareq}) appears to be too many equations. 
Because $\{{\vec r}^{\,\alpha\beta}\}$ satisfy the linear relation $\sum_{\beta}\,{\vec r}^{\,\alpha\beta}~=~0$ from Eq.(\ref{conpoint0}) there are many internal identities and
careful counting shows that it can only contain $Nd$ equations. For example if when solved Eq. (\ref{zetastareq}) were to give

\begin{equation}
\begin{aligned}
\zeta^{\,\alpha\,*}_{11}\,+\zeta^{\,\alpha\,*}_{22}\,\,=\,0 \\
\zeta^{\,\alpha\,*}_{12}\,\,=\,0
\end{aligned}
\end{equation}
then 

\begin{equation}
P \left \{ S_{ij}^{\,\alpha}|geometry \right \}=\int \prod_{\alpha}^{N}\,e^{i\sum_{\alpha}^{N}\,(S^{\,\alpha}_{11}\zeta^{\,\alpha\,*}_{11}\,+\,
2S^{\,\alpha}_{12}\zeta^{\,\alpha\,*}_{12}\,+\,S^{\,\alpha}_{22}\zeta^{\,\alpha\,*}_{22})}\, {\mathcal D}\zeta^{\,\alpha *}\, ,
\end{equation}
gives $S^{\,\alpha}_{11}~=~S^{\,\alpha}_{22}$ and no constraint on $S^{\,\alpha}_{12}\,$.
One can force Eq.(\ref{zetastareq}) into only two equations (for $d=2$) by taking scalar product with vectors $a^{\,\alpha\beta}_{i}$ and $b^{\,\alpha\beta}_{i}$ and then summing over 
$\beta$:
\begin{equation}
\sum_{\beta}\,a^{\,\alpha\beta}_{i}\,\zeta^{\,\alpha\,*}_{ij}\,r^{\,\alpha\beta}_{j}\,-\,\sum_{\beta}\,a^{\,\alpha\beta}_{i}\,\zeta^{\,\beta\,*}_{ij}\,r^{\,\beta\alpha}_{j}\,=\,0 \, , 
\label{zetastareq2}
\end{equation}

\begin{equation}
\sum_{\beta}\,b^{\,\alpha\beta}_{i}\,\zeta^{\,\alpha\,*}_{ij}\,r^{\,\alpha\beta}_{j}\,-\,\sum_{\beta}\,b^{\,\alpha\beta}_{i}\,\zeta^{\,\beta\,*}_{ij}\,r^{\,\beta\alpha}_{j}\,=\,0 \, , 
\label{zetastareq3}
\end{equation}
which gives us $2N$ equations. Suppose that we regard the differences between $\zeta^{\,\alpha\,*}_{ij}$ and $\zeta^{\,\beta\,*}_{ij}$ to be expandable in ${\vec R}^{\,\alpha\beta}$. 

\begin{equation}
\zeta^{\,\beta\,*}_{ij}\,=\,\zeta^{\,\alpha\,*}_{ij}\,+\,R^{\,\alpha\beta}_{k} \, \frac{\partial \zeta^{\,\alpha\,*}_{ij}}{\partial R^{\,\alpha}_{k}}\,+\,\dots \, .
\end{equation}
The first approximation is then

\begin{equation}
\sum_{\beta}\,a^{\,\alpha\beta}_{i}\,\zeta^{\,\alpha\,*}_{ij}\,R^{\,\alpha\beta}_{j}\,=\,0 \, ,
\label{zetastareq4}
\end{equation}
and similarly from Eq.(\ref{zetastareq3})
 
\begin{equation}
\sum_{\beta}\,b^{\,\alpha\beta}_{i}\,\zeta^{\,\alpha\,*}_{ij}\,R^{\,\alpha\beta}_{j}\,=\,0 \, .
\label{zetastareq4a}
\end{equation}
We have declared in Ref.\cite{EG} that there are two obvious vectors to be used in Eqs.(\ref{zetastareq4}) and (\ref{zetastareq4a}), namely ${\vec R}^{\,\alpha\beta}$ and 
${\vec Q}^{\,\alpha\beta}$. Hence we obtain two configuration tensors (analogous, but different from the fabric tensors used in the soil mechanics literature
\cite{Oda89}) 

\begin{equation}
F_{ij}^{\,\alpha}\,=\,\sum_{\beta}\,R^{\,\alpha\beta}_{i}\,R^{\,\alpha\beta}_{j} \, ,
\label{ct1}
\end{equation}
and
\begin{equation}
G_{ij}^{\,\alpha}\,=\,\frac{1}{2}\Big(\sum_{\beta}\,Q^{\,\alpha\beta}_{i}\,R^{\,\alpha\beta}_{j}+Q^{\,\alpha\beta}_{j}\,R^{\,\alpha\beta}_{i}\Big)\, ,
\label{ct2}
\end{equation}

then we have 

\begin{equation}
\zeta^{\,\alpha\,*}_{ij}\,F^{\,\alpha}_{ij}\,=\,0 \, ,
\label{zetastareq4b}
\end{equation}

\begin{equation}
\zeta^{\,\alpha\,*}_{ij}\,G^{\,\alpha}_{ij}\,=\,0 \, ,
\label{zetastareq4c}
\end{equation}
and after exponentiating Eqs.(\ref{zetastareq4b},\ref{zetastareq4c}) in $P \left \{ S_{ij}^{\,\alpha}|geometry \right \}$ and eliminating the auxilirary fields one
finds the missing stress-geometry equation

\begin{equation}
\begin{vmatrix}
S^{\,\alpha}_{11} &  F_{11}^{\,\alpha} & G_{11}^{\,\alpha}    \\
\\
S^{\,\alpha}_{12} &  F_{12}^{\,\alpha} & G_{12}^{\,\alpha}    \\
\\
S^{\,\alpha}_{22} &  F_{22}^{\,\alpha} & G_{22}^{\,\alpha} 
\end{vmatrix}=0\, .
\label{sgedd}
\end{equation}
The first approximation (\ref{zetastareq4},\ref{zetastareq4a}) (christened the "first coordination shell approximation") can be illustrated in the following way: we rearrange
Eq.(\ref{zetastareq}) and after multiplying by ${\vec R}^{\,\alpha\beta}$ and ${\vec Q}^{\,\alpha\beta}$ obtain 

\begin{equation}
\zeta^{\,\alpha\,*}_{ij}\,\sum_{\beta}\,R^{\,\alpha\beta}_{i}\,R^{\,\alpha\beta}_{j}\,+\,\sum_{\beta}(\zeta^{\,\alpha\,*}_{ij}-\,\zeta^{\,\beta\,*}_{ij})\,r^{\,\beta\alpha}_{j}\,R^{\,\alpha\beta}_{i}=0 \, , 
\label{zetastareq5}
\end{equation}

\begin{equation}
\zeta^{\,\alpha\,*}_{ij}\,\sum_{\beta}\,Q^{\,\alpha\beta}_{i}\,R^{\,\alpha\beta}_{j}\,+\,\sum_{\beta}(\zeta^{\,\alpha\,*}_{ij}-\,\zeta^{\,\beta\,*}_{ij})\,r^{\,\beta\alpha}_{j}\,Q^{\,\alpha\beta}_{i}=0\, ,
\label{zetastareq6}
\end{equation}
It is clear that if the second term in Eqs.(\ref{zetastareq5}) and (\ref{zetastareq6}) is neglected one obtains Eqs.(\ref{zetastareq4b}) and (\ref{zetastareq4c}).
A naive attempt to transform (\ref{sgedd}) into the macroscopic equation for $\sigma_{ij}({\vec r})$ by averaging 
$S^{\,\alpha}_{ij}$, $F^{\,\alpha}_{ij}$ and $G^{\,\alpha}_{ij}$ fails (in  the case of an isotropic configuration) because

\begin{equation}
\langle \,G^{\,\alpha}_{ij} \, \rangle \,=\, \frac{1}{V}\,\sum_{\alpha=1}^{N}\,\sum_{\beta}\,Q^{\,\alpha\beta}_{i}\,R^{\,\alpha\beta}_{j}\,=\,-
\frac{1}{V}\,\sum_{\beta=1}^{N}\,\sum_{\alpha}\,Q^{\,\beta\alpha}_{i}\,R^{\,\beta\alpha}_{j}\,=\,0
\label{ctgav}
\end{equation}
Thus for an isotropic packing the first term in Eq. (\ref{sgeomeq}) vanishes in the first coordination shell approximation. This gives rise to the conditional probability
distribution functions. Thus if we are given $S_{12}^{\,\alpha}$, the probability of finding $S_{11}^{\,\alpha}\,-\,S_{22}^{\,\alpha}$ is
\begin{equation}
P \left \{S_{11}^{\,\alpha}\,-\,S_{22}^{\,\alpha}\,|\,S_{12}^{\,\alpha}\right \}\,=\,\frac{2}{\pi}\frac{|S_{12}^{\,\alpha}|}{(S_{11}^{\,\alpha}\,-\,S_{22}^{\,\alpha})^{2}\,+\,(S_{12}^{\,\alpha})^{2}}\,.
\end{equation}
and vice versa. This distribution can then be introduced for corresponding components of the macroscopic stress tensor subject to the absence of mesoscopic
correlations in the packing.  
For an anisotropic configuration (\ref{sgedd}) characterized by the distribution of $\{G^{\,\alpha}_{ij}\}$ with some nonvanishing average yields macroscopic equation when 

\begin{equation}
\sigma_{11}\,-\,\sigma_{22}\,=\,2\sigma_{12}\,\mbox{tan} \phi \,.
\label{FPA}
\end{equation}
where $\phi$ is the angle of repose in the case when configuration is prepared in the form of a sandpile \cite{Wittmer}. 
The obvious criticism of the derivation method of Ref.\cite{EG} is that one could employ some different vector to obtain Eqs.(\ref{zetastareq4b}) and (\ref{zetastareq4c}).
For example  instead of ${\vec R}^{\,\alpha\beta}$ one could use ${\vec R}^{\,\alpha\beta}\kappa({\vec R}^{\,\alpha\beta},{\vec Q}^{\,\alpha\beta})$ where $\kappa$ is any scalar
function of ${\vec R}^{\,\alpha\beta}$ and ${\vec Q}^{\,\alpha\beta}$; or one could go to the next coordination shell of the reference grain $\alpha$ and employ 
${\vec R}^{\,\beta\gamma}$ and  ${\vec Q}^{\,\beta\gamma}$ (where $\gamma$'s are the other two neighbours of $\beta$, see Figure \ref{netiteration}) and so on. Thus we have offered a
path to the missing equation which works also in 3-D, but it is not unique \cite{Grinev}. Presumably the internal symmetries of (\ref{zetastareq}) will lead to the same
macroscopic equation when Eqs.(\ref{zetastareq4},\ref{zetastareq4a}) are used for any $a^{\,\alpha\beta}_{i}$, $b^{\,\alpha\beta}_{i}$, but it is not easy to see how. 
In the next section we will offer some new viewpoints which approach the problem from a different standpoint, but which will confirm the earlier results, and
suggest new approaches.

\section{New methods of derivation}

In the previous section we have shown that it is possible to go from Newton's equations (\ref{newt2}-\ref{couple}) to equations for conjugate 
fields (\ref{confieldeqn}) that can be then used to derive the complete set of equations for $\{S^{\,\alpha}_{ij}\}$. We can reverse the process and from 

\begin{equation}
P \left \{ S_{ij}^{\,\alpha}|geometry \right \}=\int \prod_{\alpha,\beta}^{N}\,e^{i\sum_{\alpha}^{N}\,S^{\,\alpha}_{ij}\zeta^{\,\alpha *}_{ij}}\, 
\delta(\zeta^{\,\alpha\,*}_{ij}r^{\,\alpha\beta}_{j}\,-\,\zeta^{\,\beta\,*}_{ij}r^{\,\beta\alpha}_{j})\,{\mathcal D}\zeta^{\,\alpha *}\, ,
\end{equation}
obtain  
\begin{equation}
P \left \{ S_{ij}^{\,\alpha}|geometry \right \} \,=\, \int
\,\prod_{\alpha,\beta}\,\delta(S_{ij}^{\,\alpha}-\sum_{\beta}\,p^{\,\alpha\beta}_{i}r^{\,\alpha\beta}_{j})\,
\delta(p^{\,\alpha\beta}_{i}+p^{\,\beta\alpha}_{i})\,{\mathcal D}{\vec p}^{\,\alpha\beta}\, .
\label{stresspdfstar3}
\end{equation}
Since the only constraint on $\left \{{\vec p}^{\,\alpha\beta}\right \}$ is given by
\begin{equation}
p^{\,\alpha\beta}_{i}+p^{\,\beta\alpha}_{i}\,=\,0
\label{pfnewt3}
\end{equation}
the missing stress-geometry equation comes from the condition on $S_{ij}^{\,\alpha}$ that 
\begin{equation}
S_{ij}^{\,\alpha}\,=\,\frac{1}{2}\sum_{\beta}\,\Big(p^{\,\alpha\beta}_{i}r^{\,\alpha\beta}_{j}\,+\,p^{\,\alpha\beta}_{j}r^{\,\alpha\beta}_{i}\Big)
\label{sdefpf}
\end{equation}
can be satisfied by a set of "pseudo-forces" $\{{\vec p}^{\,\alpha\beta}\}$ which obey (\ref{pfnewt3}), but are free from the constraint of Newton's second law
(\ref{newt2}).  "Pseudo-force"  ${\vec p}^{\,\alpha\beta}$ can be written in the following form
\begin{equation}
p_{i}^{\,\alpha\beta}\,=\,(\psi^{\,\alpha} \,+\, \psi^{\,\beta})\,R_{i}^{\,\alpha\beta}\,+\,(\chi^{\,\alpha} \,-\, \chi^{\,\beta})\,Q_{i}^{\,\alpha\beta} \, ,
\end{equation}
which satisfies (\ref{pfnewt3}) and provides Eq. (\ref{sgedd}).
We can think of $\{S^{\,\alpha}_{ij}\}$ as a $\frac{Nd(d+1)}{2}$ component vector ${\vec S}$, and $\{{\vec p}^{\,\alpha\beta}\}$ as 
a $\frac{Nd(d+1)}{2}$ component vector ${\vec P}$. Then the set of equations (\ref{sdefpf}) can be written as
\begin{equation}
{\vec S}\,=\,{\vec R}\,{\vec P}
\label{mat}
\end{equation}
where ${\vec R}$ is a $\frac{Nd(d+1)}{2}\,\times\,\frac{Nd(d+1)}{2}$ matrix with a large repetition of elements since $\sum_{\beta}\,{\vec r}^{\,\alpha\beta}~=~0$.
This matrix has $\frac{Nd(d-1)}{2}$ zero eigenvalues which gives the right number of constraints on $\{S^{\,\alpha}_{ij}\}$. The argument here is of the 
"must be so" type. Apart from the easy proof that $\mbox{Det}\,{\vec R}~=~0$ (by adding rows) we have not succeeded in proving that the
$\frac{Nd(d+1)}{2}\,\times\,\frac{Nd(d+1)}{2}$ matrix has $\frac{Nd(d-1)}{2}$ zero eigenvalues, but it "must be so"!
Another and perhaps easier method is to return to the Eq.(\ref{zetastareq}) for $\zeta^{\,\alpha\,*}_{ij}$ and sum it over $\beta$ using 
$\sum_{\beta}\,\vec{r}^{\,\alpha\beta}~=~0$. Therefore we have

\begin{equation}
\sum_{\beta}\,\zeta^{\,\beta\,*}_{ij}r^{\,\beta\alpha}_{j}\,=\,0 \, .
\end{equation}
This can be used in Eq.(\ref{stresspdfstar3})
\begin{equation}
P \left \{ S_{ij}^{\,\alpha}|geometry \right \}=\int \prod_{\alpha,\beta}^{N}\,e^{i\sum_{\alpha}^{N}\,S^{\,\alpha}_{ij}\zeta^{\,\alpha *}_{ij}}\, 
\delta(\sum_{\beta}\,\zeta^{\,\beta\,*}_{ij}r^{\,\beta\alpha}_{j})\,{\mathcal D}\zeta^{\,\alpha *}\, ,
\end{equation}
as in the derivation of (\ref{sdefpf}), but now losing a positional index
\begin{equation}
S^{\,\alpha}_{ij}\,=\,\frac{1}{2}\sum_{\beta}\,\Big(\phi^{\,\beta}_{i}\,r^{\,\alpha\beta}_{j}\,+\,\phi^{\,\beta}_{j}\,r^{\,\alpha\beta}_{i}\Big)\, .
\label{microstress2}
\end{equation}
As before this equation implies a relationship between the components of $S^{\,\alpha}_{ij}$. Suppose now that we are looking for an ansatz for 
$\phi^{\,\beta}_{i}$. Since $\phi^{\,\beta}_{i}$ is a vector it can be represented as a superposition of two obvious candidates ${\vec R}^{\,\alpha\beta}$ and 
${\vec Q}^{\,\alpha\beta}$
\begin{equation}
\phi_{i}^{\,\beta}\,=\,\psi^{\,\alpha}\,R_{i}^{\,\alpha\beta}\,+\,\chi^{\,\alpha}\,Q_{i}^{\,\alpha\beta}
\label{Airy}
\end{equation}
where new quantities $\psi^{\,\alpha}$ and $\chi^{\,\alpha}$ are scalars. This gives us
\begin{equation}
S^{\,\alpha}_{ij}\,=\,\psi^{\,\alpha}\Big(F^{\,\alpha}_{ij}\,+\,G^{\,\alpha}_{ij}\Big)\,+\,\chi^{\,\alpha}\Big(H^{\,\alpha}_{ij}\,+\,G^{\,\alpha}_{ij}\Big) \, ,
\end{equation}
where $H^{\,\alpha}_{ij}~=~\sum_{\beta}\,Q_{i}^{\,\alpha\beta} Q_{j}^{\,\alpha\beta}$.
Eliminating $\psi^{\,\alpha}$ and $\chi^{\,\alpha}$ leads to

\begin{equation}
\begin{vmatrix}
S^{\,\alpha}_{11} &  F_{11}^{\,\alpha}+G_{11}^{\,\alpha} & H_{11}^{\,\alpha}+G_{11}^{\,\alpha}   \\
\\
S^{\,\alpha}_{12} &  F_{12}^{\,\alpha}+G_{12}^{\,\alpha} & H_{12}^{\,\alpha}+G_{12}^{\,\alpha}    \\
\\
S^{\,\alpha}_{22} &  F_{22}^{\,\alpha}+G_{22}^{\,\alpha} & H_{22}^{\,\alpha}+G_{22}^{\,\alpha} 
\end{vmatrix}=0\, .
\label{sgedd2}
\end{equation}
Note that $H_{ij}^{\,\alpha}$ will on average be a multiple of $F_{ij}^{\,\alpha}$ and this gives us Eq. (\ref{sgedd}).
It is possible to obtain from (\ref{microstress2})the third term in (\ref{sgeomeq}). The second term is non-vanishing only in special cases of periodic arrays
\cite{Ball}. Let us construct the following interpolation of $\phi^{\,\beta}_{i}$
\begin{equation}
\phi_{i}^{\,\beta}\,=\,\phi_{i}^{\,\alpha}\,+\,R_{j}^{\,\alpha\beta}\,\nabla_{j}\,\phi_{i}^{\,\alpha} \, ,
\end{equation}
after substituiting it into (\ref{microstress2}) and summing it over $\beta$ using 
$\sum_{\beta}\,\vec{r}^{\,\alpha\beta}~=~0$ we have
\begin{equation}
S^{\,\alpha}_{ij}\,=\,\nabla_{k}\,\phi_{i}^{\,\alpha}\,\sum_{\beta}\,R_{k}^{\,\alpha\beta}\,r_{j}^{\,\alpha\beta} \, .
\end{equation}
Crude averaging gives us
\begin{equation}
\sigma_{ij}=\frac{1}{V}\,\sum_{\alpha}^{N}\,\sum_{\beta}\,r_{j}^{\,\alpha\beta}\,R_{k}^{\,\alpha\beta}\,\nabla_{k}\,\phi_{i}^{\,\alpha}\,=\,F_{jk}\nabla_{k}\,\phi_{i} \, .
\label{macrostress2}
\end{equation}
We can now eliminate $\phi_{i} $ and obtain the well-known Navier equation which imposes kinematic compatibility on the stresses \cite{Flugge} 

\begin{equation}
\frac{\partial^{2} \sigma_{xx}}{\partial y^{2}}\,+\,\frac{\partial^{2}
\sigma_{yy}}{\partial x^{2}}\,-\,2\frac{\partial^{2} \sigma_{xy}}{\partial x \partial
y}\,=\,0\, .
\label{Navier}
\end{equation}
This equation corresponds to the third term in (\ref{sgeomeq}) in the case of an isotropic packing and implies that the stress tensor components can be expressed
in terms of the Airy function \cite{Flugge} whose discrete analogues are  $\psi^{\,\alpha}$ and $\chi^{\,\alpha}$ in (\ref{Airy}).This also means that the set of
microscopic constraints (\ref{microstress2}) is consistent with the macroscopic equation (\ref{sfeqmacro}).
\begin{figure}[h] 
\begin{center} 
\resizebox{6cm}{!}{\rotatebox{0}{\includegraphics{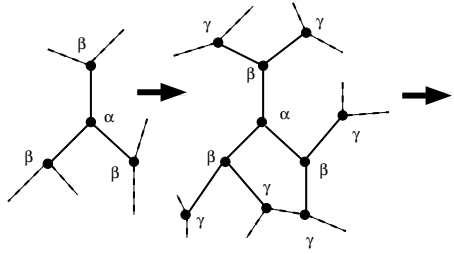}}}
\caption{Propagation of coordination shell vectors through the contact network}
\label{netiteration} 
\end{center} 
\end{figure}
However in 3-D we have encountered mathematical difficulties with this simple averaging procedure.
This highlights the puzzle we face. One could put much more complicated versions of $\phi^{\,\beta}_{i}$, for example any functional form which employs vectors constructed out
of the vectors of the Figure \ref{centroid} or indeed the next coordination shell (see Figure \ref{netiteration}). The problem is a geometric puzzle. A set of contact points $\{{\vec{R}}^{\,\alpha\beta *}\}$
corresponds to the packing of grains with coordination number $z=d+1$. From these points associated with the reference grain $\alpha$ one can construct the vectors of Figure 
\ref{centroid}, and find the basic equations using $\zeta^{\,\alpha\,*}_{ij}$ or ${\vec p}^{\,\alpha\beta}$, or $\phi^{\,\alpha}_{i}$. With these equations one may take {\em ad
hoc} steps to obtain the complete set of equations for the macroscopic stress tensor $\sigma_{ij}({\vec r})$, but we have failed to find a systematic procedure such as is
possible for the corrections of the stress-force equation as outlined in Ref.\cite{EG}. The problem is now purely geometric, but we emphasize that although real granular materials
have many features omitted here, we are studying  the simplest possible case and the geometric puzzle offered here although difficult is quite basic, and no simpler case can be
found.
\section{Acknowledgments}

We wish to acknowledge the financial support of Leverhulme Foundation S.F.E), ROPA grant from EPSRC (UK) and Research Fellowship from Wolfson College (D. V. G.).
Authors thank Prof. R. C. Ball and Dr. R. Blumenfeld (who have derived a different route to solve this problem \cite{Ball00}) for stimulating discussions.


\begin{thebibliography}{99}
\bibitem{Jaeger}{\it Granular Matter: An Interdisciplinary Approach}, A. Mehta (Ed.) (Springer-Verlag, New-York, 1993), 
for review see e.g.  H. M. Jaeger, S. R. Nagel, and R. P. Behringer, Rev.Mod.Phys. {\bf 68}, 1259 (1996).

\bibitem{EG} S. F. Edwards and D. V. Grinev, Phys. Rev. Lett., {\bf 82}, 5397 (1999),

\bibitem{Pars} L. A. Pars, A Treatise on Analytical Dynamics, Chapter XII, (Heinemann, London 1968).

\bibitem{Oda89} M.~Oda, T.~Sudoo, {\em Powders and grains}, J.~Biarez and R.~Gourves (Eds.), 155,(A. A. Balkema, Rotterdam, 1989), and references therein.

\bibitem{Wittmer} J. P. Wittmer, P. Claudin, M. E. Cates,  J. de Physique I (France) {\bf 7} , 39 (1997).

\bibitem{Ball} R. C. Ball and D. V. Grinev, Physica A {\bf 292}, 167 (2001); R. C. Ball{\em Structure and Dynamics of Materials in the Mesoscopic
Domain}, M.Lal, R. A. Mashelkar, B.D. Kulkarni, V.M.Naik (Eds.), 326,(Imperial College Press, London, 1999).

\bibitem{Flugge} {\em Encyclopedia of Physics}, v.VI, {\em Elasticity and Plasticity},  S. Fl\"{u}gge (Ed.), (Springer-Verlag, Berlin, 1958) or 
K. Washizu, {\em Variational Methods in Elasticity and Plasticity}, Pergamon Press, Oxford (1982), Chapter 1 .

\bibitem{Grinev} D. V. Grinev, {\em in preparation}.

\bibitem{Ball00} R. C. Ball and R. Blumenfeld, {\it cond-matt/0008127}. 

\end{thebibliography}
\end{document}